\documentclass{article}

\PassOptionsToPackage{numbers,compress}{natbib}
\usepackage{titletoc}
 \usepackage[preprint]{neurips_2026}

\usepackage[utf8]{inputenc} 
\usepackage[T1]{fontenc}    
\usepackage{hyperref}
\hypersetup{hidelinks}         
\usepackage{url} 
\usepackage{multirow}
\usepackage{amsmath}    
\usepackage{xcolor} 
\usepackage[table]{xcolor}
\usepackage{booktabs}       
\usepackage{amsfonts}       
\usepackage{nicefrac}       
\usepackage{microtype}      
\usepackage{xcolor}         
\usepackage{subcaption}
\usepackage{enumitem}
\usepackage{graphicx}
\usepackage{longtable}     
\usepackage{pdflscape}
\usepackage{float}

\definecolor{rowgray}{gray}{0.9}

\newcommand{\iiot}{IIoT}

\title{Cross-Domain Generalization Failure in Lightweight
  Intrusion Detection Models for IIoT Networks}

\author{%
  \textbf{MD Azizul Hakim}$^{1*}$ \quad \textbf{Md Shihab Uddin}$^{2}$ \quad \textbf{Talha Ibne Anich}$^{1}$ \\[0.4em]
  \texttt{azizulhakim8291@gmail.com} \quad
  \texttt{shihab9586@gmail.com} \quad
  \texttt{mail.talhaibneanich@gmail.com} \\[0.6em]
  $^{1,3}$Department of Computer Science and Technology,
  Bangladesh Sweden Polytechnic Institute \\
  $^{2}$Department of Electrical Engineering,
  Bangladesh Sweden Polytechnic Institute \\
  Kaptai, Rangamati, Chittagong 4530, Bangladesh \\[0.3em]
  {\small $^{*}$Corresponding author}
}
\usepackage{titlesec}
\titlespacing*{\section}{0pt}{6pt plus 2pt minus 2pt}{4pt plus 1pt}
\titlespacing*{\subsection}{0pt}{5pt plus 2pt minus 2pt}{3pt plus 1pt}
\titlespacing*{\subsubsection}{0pt}{4pt plus 1pt minus 1pt}{2pt plus 1pt}

\begin{document}
\setlength{\abovedisplayskip}{6pt}
\setlength{\belowdisplayskip}{6pt}
\setlength{\abovecaptionskip}{4pt}
\setlength{\belowcaptionskip}{-4pt}

\maketitle
\begin{abstract}
Lightweight machine learning models are increasingly proposed for
intrusion detection in Industrial Internet of Things (\iiot) networks
due to their suitability for resource-constrained edge deployment.
However, most reported results evaluate these models only within their
training network, leaving behavior on unseen industrial networks largely
unverified. This study trains four representative lightweight
architectures on one widely used \iiot{} dataset and evaluates them,
without retraining, on two independent and structurally distinct \iiot{}
datasets, using a feature representation restricted to attributes
available across all three sources. Explainability analysis, corroborated across two architecturally distinct top-performing models, shows that both rely overwhelmingly on coarse port-category features, and a direct comparison of port-category prevalence across
datasets reveals that the most influential category occurs in source-domain attack traffic at 96 to 435 times the rate observed in the two target domains, indicating that coarsening port resolution relocates rather than removes a documented shortcut.
Evaluation under naturally imbalanced class distributions, rather than
the balanced distributions common in prior work, reveals a further
effect: the evaluation protocol used can reverse which target network
appears to pose the greater generalization challenge. Adversarial
robustness and the capacity to recover performance through limited
exposure to target-domain data are also assessed; robustness to
adversarial perturbation is found to be unrelated to a model's ability
to generalize across networks, and recovery through limited adaptation
varies considerably by architecture. These findings suggest that
deployment readiness for lightweight \iiot{} intrusion detection should
be assessed using cross-network evaluation under realistic class
distributions, rather than within-domain accuracy alone.
\end{abstract}

\section{Introduction}
\label{sec:introduction}

Industrial Internet of Things (IIoT) networks now connect sensors, controllers, and actuators across manufacturing, energy, and critical infrastructure systems, creating an attack surface that conventional, server-grade intrusion detection cannot always reach. This has driven sustained interest in \emph{lightweight} machine learning models for intrusion detection: classifiers small enough to run on constrained edge hardware, where computation, memory, and energy budgets are limited. A growing body of work reports such models achieving near-perfect detection accuracy on standard IIoT traffic benchmarks, suggesting that the problem of resource-efficient intrusion detection is largely solved.

This impression rests on a narrow form of evaluation. In almost all published studies, a model is trained and tested on data drawn from the same capture, often a single split of one dataset collected on one network. Real IIoT deployments do not work this way. A model trained on one industrial network is routinely expected to operate on another, with different devices, different traffic patterns, and different attacker behavior. Whether the accuracy reported under same-dataset evaluation has any bearing on this realistic deployment scenario is rarely tested, and almost never tested for the specific class of lightweight models that edge deployment requires.

This gap motivates a direct question: \emph{do lightweight intrusion detection models that perform well within their training network retain that performance when applied, without retraining, to a different IIoT network?} A closely related question follows immediately: if performance does not transfer, is the cause identifiable, and can it be addressed cheaply, through a small amount of local adaptation, rather than full retraining?

This paper answers both questions through a systematic cross-dataset study. Four representative lightweight architectures, a decision tree~\cite{loh2011classification} and three compact neural network designs, are trained on one IIoT dataset and evaluated, without retraining, on two independent IIoT datasets that differ in source network, device composition, and traffic capture format. To isolate the cause of any performance loss, feature reliance is examined directly through explainability analysis, and a controlled ablation tests whether reducing the resolution of the most influential feature removes the problem. Evaluation is repeated under both artificially balanced and naturally imbalanced class distributions, since the former is the de facto standard in prior work but does not reflect deployment conditions. Adversarial robustness is assessed independently of cross-dataset performance, and a final experiment measures whether fine-tuning on a small, labeled sample from the target network can recover what zero-shot transfer loses.

The contribution of this work is fourfold:

\begin{itemize}
    \item It provides direct evidence, across two independent target datasets, that strong within-domain performance for lightweight IIoT intrusion detection models does not imply cross-network performance~\cite{latha2023analysis}.
    \item It shows, through explainability analysis, a targeted ablation, and a direct cross-dataset comparison of port-bucket prevalence, that coarsening port resolution into broad categorical buckets rules out raw port memorization as the cause of failure, but that reliance shifts to the coarser port-bucket categories themselves; the most influential of these categories occurs in source-domain attack traffic at approximately 96 times the rate observed in Gotham's attack traffic and 435 times the rate in WUSTL-IIoT-2021's, confirming that reducing feature granularity relocates this shortcut rather than removing it.
    \item It shows that evaluation under artificially balanced class distributions, common in prior work, substantially misrepresents performance relative to the natural, imbalanced conditions a deployed system would encounter.
    \item It demonstrates that the capacity to recover performance through small-scale fine-tuning on target-network data is architecture-dependent, rather than a property of lightweight models in general.
\end{itemize}

Together, these results indicate that the readiness of a lightweight intrusion detection model for IIoT deployment cannot be inferred from within-domain accuracy, and that cross-network evaluation should be treated as a standard requirement rather than an optional extension.
\section{Related Work}
\label{sec:related_work}

\subsection{Lightweight Intrusion Detection for IIoT}

A substantial body of work targets intrusion detection models small enough for edge deployment in IIoT settings. Ferrag et al. introduced Edge-IIoTset \cite{ferrag2022edge}, a testbed-generated dataset spanning fourteen attack categories across IoT and IIoT devices, and reported high detection accuracy using a range of classifiers, evaluated entirely within the dataset's own train/test split. Zolanvari et al. introduced the WUSTL-IIoT line of datasets through a vulnerability-analysis study of industrial IoT networks \cite{zolanvari2019machine}, which has since been used by numerous lightweight-classifier studies, again under single-dataset evaluation. Belarbi et al. released the Gotham dataset \cite{belarbi2025gotham}, a large-scale, reproducible capture of real heterogeneous IoT device traffic, intended in part to address the lack of diversity in existing IoT security benchmarks. These three datasets, used respectively as the training domain and two evaluation domains in the present study, were each introduced with in-domain results only; none of their originating publications report performance when a model trained on one is tested on another.

\subsection{Cross-Dataset Generalization in Intrusion Detection}

A smaller, more recent line of work has begun to test whether intrusion detection performance survives a change of network. Cantone et al. \cite{cantone2024machine} trained four classifiers on one enterprise network dataset and evaluated them on three independent variants, reporting near-perfect within-dataset accuracy collapsing to chance-level performance across datasets, and attributed the collapse to dataset-specific anomalies identified through visualization rather than to a single confounding feature. D'Hooge et al. \cite{d2019depth} reported a comparable failure restricted to two attack categories across two enterprise datasets, concluding that the classifiers tested could not operate reliably on traffic from a network unseen at training time. Verkerken et al. \cite{verkerken2022towards} extended this observation to unsupervised models, reporting a substantial average drop in detection performance under cross-dataset evaluation relative to within-dataset evaluation. Apruzzese, Pajola, and Conti~\cite{apruzzese2022cross} proposed XeNIDS, a structured framework for cross-evaluating ML-based network intrusion detection systems across multiple datasets, showing that cross-evaluation can both extend a model's effective detection surface at no additional labeling cost and expose risks that single-dataset evaluation conceals. Layeghy and Portmann~\cite{layeghy2023explainable} combined explainability analysis with cross-domain NIDS evaluation, the closest existing work to the present study in spirit; their analysis identifies feature-importance shifts across datasets but does not test a targeted ablation isolating a single confounding feature, compare balanced against natural class distributions, or assess few-shot recovery. These studies establish that cross-dataset failure is a recurring property of network intrusion detection generally; none of them isolate lightweight architectures specifically, and none test IIoT datasets, which differ from the enterprise network traffic these studies examine in device composition, protocol mix, and traffic volume.

A second, closely related line of work has identified the specific mechanism most often blamed for inflated intrusion detection performance: reliance on fields that identify a flow's origin rather than its behavior. Engelen et al. \cite{engelen2021troubleshooting} showed that classifiers trained on a widely used benchmark dataset could achieve high accuracy using destination port alone, demonstrating that apparent detection performance can reflect a learned association between an attack's source and a fixed port number rather than any property of malicious traffic itself. This finding has informed subsequent dataset-design work; a recent study on IoT intrusion detection explicitly removed IP- and port-based identifiers from its feature set to obtain a more behavior-driven, and presumably more transferable, representation \cite{dharini2026efficient}. The present study tests this mechanism directly, rather than designing around it: feature reliance is measured before and after reducing port resolution, and cross-dataset performance is compared under both conditions, allowing the contribution of this specific, previously documented shortcut to be isolated from whatever remains.

A small number of more recent studies have moved cross-dataset evaluation specifically into the IIoT/IoT setting. Bilal et al. \cite{bilal2026dataset} evaluated federated intrusion detection models across three IIoT datasets, including Edge-IIoTset, reporting macro-F1 losses of up to thirty percentage points when a model trained on one dataset is tested on another under a single-source, non-federated condition. This study excludes IP addresses and port identifiers from its feature set entirely by design, so it does not test whether port reliance contributes to the observed transfer loss, and its architectures are evaluated within a federated aggregation process rather than in isolation, which prevents attributing generalization behavior to any single lightweight model. This confirms that cross-dataset failure extends to IIoT data specifically, strengthening rather than undermining the motivation for the present work; it does not, however, isolate architecture-level behavior, test the port-shortcut hypothesis directly, compare balanced against naturally imbalanced evaluation, or assess few-shot adaptation.

\subsection{Adversarial Robustness of Intrusion Detection Models}

Separately from generalization, the robustness of intrusion detection classifiers to adversarially perturbed input has been studied as a deployment risk in its own right. Vitorino et al. \cite{vitorino2023towards} examined the vulnerability of IoT intrusion detection models, including tree-based classifiers, to realistic adversarial perturbation, reporting substantial degradation under attack. Debicha et al. \cite{debicha2023review} surveyed evasion attacks and defenses for network intrusion detection broadly, noting that robustness is rarely evaluated alongside generalization in the same study. The present work follows this observation by assessing adversarial robustness independently from, but alongside, cross-dataset performance for each architecture.

\subsection{Domain Adaptation and Few-Shot Recovery}

Where generalization failure has been observed, a parallel body of work has examined whether it can be mitigated without full retraining. Singla et al. \cite{singla2020preparing} used adversarial domain adaptation to prepare intrusion detection models with minimal labeled target-domain data, reporting recovered performance using a small fraction of the data otherwise required. Kheddar et al. \cite{kheddar2023deep} reviewed deep transfer learning approaches for intrusion detection in industrial control networks, covering a range of adaptation strategies but noting limited comparison of how adaptability varies across lightweight model architectures specifically. The present study addresses this directly by applying the same fine-tuning procedure, under matched conditions, to all four evaluated architectures.

\subsection{Summary of the Gap}

Across these bodies of work, no study combines (i) lightweight IIoT-specific architectures evaluated in isolation, (ii) more than one independent target dataset, (iii) a direct test of the port-shortcut mechanism rather than an assumption that it explains the failure or a design choice to remove it by default, (iv) a comparison of evaluation under balanced versus naturally imbalanced class distributions, and (v) a matched few-shot adaptation test across all evaluated architectures, as summarized in Table~\ref{tab:gap_table}. This study is constructed to close that combined gap rather than any single piece of it in isolation.
\begin{table}[t]
\centering
\small
\caption{Coverage of related cross-dataset IDS studies against the five criteria this paper combines: (i) lightweight architectures evaluated in isolation, (ii) more than one independent target dataset, (iii) a direct test of the port-shortcut mechanism, (iv) balanced vs.\ natural class-distribution comparison, (v) matched few-shot adaptation across architectures.}
\label{tab:gap_table}
\begin{tabular}{lccccc}
\toprule
\textbf{Study} & \textbf{(i)} & \textbf{(ii)} & \textbf{(iii)} & \textbf{(iv)} & \textbf{(v)} \\
\midrule
Cantone et al.~(\citeyear{cantone2024machine})          & -- & \checkmark & -- & -- & -- \\
D'Hooge et al.~(\citeyear{d2019depth})                  & -- & -- & -- & -- & -- \\
Verkerken et al.~(\citeyear{verkerken2022towards})      & -- & -- & -- & -- & -- \\
Apruzzese et al.~(\citeyear{apruzzese2022cross})              & -- & \checkmark & -- & -- & -- \\
Layeghy \& Portmann~(\citeyear{layeghy2023explainable}) & -- & \checkmark & -- & -- & -- \\
Bilal et al.~(\citeyear{bilal2026dataset})               & -- & \checkmark & -- & -- & -- \\
\textbf{This work}                                       & \checkmark & \checkmark & \checkmark & \checkmark & \checkmark \\
\bottomrule
\end{tabular}
\end{table}
\section{Datasets and Feature Engineering}
\label{sec:datasets}

Three IIoT/IoT network traffic datasets are used: \textbf{Edge-IIoTset} \cite{ferrag2022edge}, the training domain, comprising fourteen attack categories across IoT and IIoT testbed devices; \textbf{Gotham 2025} \cite{belarbi2025gotham}, a reproducible capture of real per-device IoT traffic across 104 devices, used as the first cross-domain target; and \textbf{WUSTL-IIoT-2021} \cite{ahuja2021wustl}, an Argus flow-export capture from an industrial SCADA testbed, used as the second cross-domain target. The three datasets were collected with different tools, at different layers (packet versus flow), and expose different native fields, so no raw feature is available in all three.

A minimal \emph{common schema} was constructed instead: source/destination port, transport protocol, and four TCP flag bits (FIN, SYN, RST, ACK). Raw port numbers were not used directly. Prior work has shown that destination port alone can yield high detection accuracy on benchmark intrusion datasets \cite{engelen2021troubleshooting}, because port values often identify a flow's dataset of origin rather than its behavior. To prevent this shortcut and obtain a feature representation with a chance of transferring across networks, each port is instead mapped to one of four categorical buckets, \emph{none}, \emph{well-known} ($<1024$), \emph{registered} ($1024$--$49151$), or \emph{dynamic} ($\geq 49152$), and one-hot encoded~\cite{seger2018investigation} separately for source and destination. Protocol is one-hot encoded into TCP, UDP, ICMP, or OTHER, the last capturing any protocol value not matching the first three (e.g., GRE, ESP, or unparsed/missing protocol fields). This yields a 16-dimensional schema (4 flag bits + 8 port-bucket indicators + 4 protocol indicators), identical across all three datasets despite their different native formats. WUSTL-IIoT-2021's Argus export exposes no TCP flag information; the four flag features are zero-filled for this dataset, a disclosed limitation rather than a missing-data error.

\begin{table}[h]
\centering
\caption{Dataset summary. Source/target role, size after capping, and class balance.}
\label{tab:datasets}
\resizebox{\columnwidth}{!}{%
\begin{tabular}{lccc}
\toprule
\rowcolor{rowgray}
& \textbf{Edge-IIoTset} & \textbf{Gotham 2025} & \textbf{WUSTL-IIoT-2021} \\
\midrule
Role & Source (train) & Target 1 & Target 2 \\
Capture type & Packet-level & Packet-level & Flow-level (Argus) \\
Devices/files & 1 & 104 & 1 \\
$n$ (balanced eval) & 299{,}461 & 118{,}444 & 236{,}959 \\
$n$ (natural eval) & --- & 189{,}396 & 299{,}832 \\
Attack share (natural) & --- & 10.6\% & 7.3\% \\
TCP flag bits available & Yes & Yes & \cellcolor{rowgray}No \\
\bottomrule
\end{tabular}}
\end{table}

Attack labels are harmonized to a binary attack/benign distinction across all three datasets despite differing native taxonomies: Edge-IIoTset's fourteen named attack categories, Gotham's per-device anomaly labels, and WUSTL-IIoT-2021's binary \texttt{Target} field are each mapped to a single binary label, discarding multiclass distinctions for the cross-domain experiments (Section~\ref{sec:results_feature}'s multiclass analysis uses Edge-IIoTset's native categories directly, in-domain only)
All models are trained once on Edge-IIoTset and evaluated, without retraining, on Gotham and WUSTL-IIoT-2021. Each target dataset is evaluated twice: once under class-balanced sampling, the convention in most prior cross-dataset studies, and once under its natural, heavily benign-skewed class distribution, to test whether balanced evaluation overstates real-world performance. Table~\ref{tab:datasets} summarizes role, size, and class balance for all three.

\section{Methodology}
\label{sec:methodology}

\subsection{Model Architectures}

Four lightweight classifiers are trained on the 16-dimensional common schema (Section~\ref{sec:datasets}): \textbf{DecisionTree}~\cite{loh2011classification}, a depth-limited tree ($\text{max\_depth}=10$); \textbf{SmallMLP}~\cite{taud2017multilayer}, a three-layer feed-forward network ($16 \rightarrow 16 \rightarrow 8 \rightarrow 2$, ReLU); \textbf{Small1DCNN}~\cite{ige2024state}, two 1D convolutional layers (8 and 16 channels, kernel size 3) followed by global average pooling and a linear classifier, with no normalization layers; and \textbf{SmallLSTM}~\cite{graves2012long}, a single-layer LSTM (hidden size 16) reading each feature vector as a length-16 sequence of scalars, followed by a linear classifier on the final hidden state. This treats the 16-dimensional feature vector as a sequence purely to obtain a lightweight recurrent baseline; no temporal ordering between flows is assumed or exploited, and the choice trades a small increase in parameter count (Table~\ref{tab:results}, Section~\ref{sec:results_efficiency}) for a recurrent inductive bias as a point of comparison against the feed-forward and convolutional architectures. All three neural architectures are trained with the Adam optimizer minimizing class-weighted cross-entropy, using a learning rate of $1\times10^{-3}$, batch size 256, for 15 epochs.
\begin{equation}
\mathcal{L} = -\sum_{i=1}^{N} w_{y_i} \log \hat{p}_{i,y_i},
\label{eq:loss}
\end{equation}
where $\hat{p}_{i,y_i}$ is the predicted probability of the true class $y_i$ for sample $i$, and $w_{y_i}$ is the inverse class frequency, set to 1 for all classes during initial training and re-computed per fine-tuning batch during few-shot adaptation (Section~\ref{sec:fewshot}).

\subsection{In-Domain and Cross-Domain Evaluation}

Edge-IIoTset is split 70/10/20 into train, validation, and test sets, stratified by label. Each model is trained once on the training split and evaluated on the held-out test split (in-domain). The same trained models, without any further parameter updates, are then evaluated on Gotham and on WUSTL-IIoT-2021 (cross-domain), each under both a class-balanced sample and the dataset's natural class distribution (Table~\ref{tab:datasets}). All four conditions use the same \texttt{StandardScaler} (zero mean, unit variance), fit once on Edge-IIoTset's training split. Detection performance is reported as the F1 score~\cite{van1979information} on the attack class,
\begin{equation}
F_1 = \frac{2 \cdot \text{Precision} \cdot \text{Recall}}{\text{Precision} + \text{Recall}},
\label{eq:f1}
\end{equation}
which, unlike accuracy, does not reward a classifier for defaulting to the majority (benign) class under the natural, imbalanced evaluation condition.

\subsection{Efficiency Profiling}

For each trained model, four lightweight-deployment metrics are recorded: serialized model size on disk, single-sample inference latency (mean, median, 95th and 99th percentile over 1{,}000 repeated calls), peak memory during inference, and parameter count (node count for DecisionTree, learnable parameter count for the neural models). Training wall-clock time is recorded separately for each architecture under identical hardware conditions.

\subsection{Explainability}

To diagnose what each model relies on, SHAP~\cite{lundberg2017unified} values are computed for the best in-domain model (by F1) over a fixed sample of 200 test instances: \texttt{TreeExplainer} for DecisionTree, \texttt{KernelExplainer} for the three neural models. Mean absolute SHAP value per feature, averaged over the sample, ranks feature contribution to the attack-class prediction. Decision-tree feature importances (mean decrease in impurity) are reported as a second, model-native measure of feature reliance, to cross-check the SHAP ranking.

\subsection{Adversarial Robustness}

Robustness to evasion is tested with the HopSkipJump black-box attack \cite{chen2020hopskipjumpattack,nicolae2018adversarial}, applied identically to all four models (including DecisionTree, which gradient-based attacks such as FGSM cannot target). A fixed random sample of 100 in-domain test instances is perturbed ($\text{max\_iter}=20$, $\text{max\_eval}=500$, $\text{init\_eval}=100$), and robustness is reported as the accuracy drop,
\begin{equation}
\Delta\text{Acc} = \text{Acc}_{\text{clean}} - \text{Acc}_{\text{adv}},
\label{eq:advdrop}
\end{equation}
between the model's accuracy on the clean sample and on its adversarially perturbed counterpart.

\subsection{Statistical Validation}

Two checks establish that the reported differences are not artifacts of a single train/test split. First, each architecture is retrained from scratch across five random seeds, and in-domain and cross-domain F1 are reported as mean $\pm$ standard deviation. Second, McNemar's test~\cite{pembury2020effective} is applied pairwise between all six model pairs, on identical test instances, separately for the in-domain condition and the cross-domain condition on Gotham under balanced evaluation (the same condition used for multi-seed validation, Section~\ref{sec:results_validation}):
\begin{equation}
\chi^2 = \frac{(|b - c| - 1)^2}{b + c},
\label{eq:mcnemar}
\end{equation}
where $b$ and $c$ are the off-diagonal counts of the $2\times2$ contingency table of correct/incorrect predictions for the two models being compared (continuity-corrected). A pair is reported as significantly different at $p < 0.05$.

\subsection{Few-Shot Domain Recovery}
\label{sec:fewshot}

To test whether the cross-domain gap can be closed cheaply, Gotham is split in half into a fixed adaptation pool and a fixed evaluation set, held constant across all conditions. For fractions $\{0, 0.01, 0.05, 0.10, 0.25\}$ of the adaptation pool, each model is updated using only that fraction of labeled target-domain data: DecisionTree~\cite{loh2011classification} is refit on the union of its original training data and the sampled fraction; the three neural models are fine-tuned for 5 additional epochs at one-tenth the original learning rate, with class weights recomputed on the fine-tuning batch (Eq.~\ref{eq:loss}) to counter the class imbalance typical of small samples. This procedural difference is intentional, reflecting how each model type is conventionally updated in practice, but it means DecisionTree's recovery reflects retraining on its full original training set plus the new sample, while the neural models' recovery reflects fine-tuning on the new sample alone; the two are not directly comparable as a controlled test of architecture alone. F1 on the fixed evaluation set (Eq.~\ref{eq:f1}) is tracked across all five fractions, including the $\text{fraction}=0$ zero-shot baseline, for a direct, fraction-by-fraction recovery curve.

\section{Results}
\label{sec:results}
Results are organized to separate what each model achieves from why. Section~\ref{sec:results_indomain} establishes that all four architectures fit the training domain equally well, removing architecture choice as a source of the differences that follow. Section~\ref{sec:results_crossdomain} reports the central finding: detection performance collapses when each model, without retraining, is applied to traffic from a different IIoT network, and this collapse holds under two independent target datasets and under both balanced and naturally imbalanced evaluation. Sections~\ref{sec:results_validation} through~\ref{sec:results_sizecompare} then test, in turn, whether this collapse is statistically reliable, what each model relies on to make its decisions, how robust each model is to adversarial perturbation, whether the gap can be closed with a small amount of target-domain data, what each architecture costs to deploy, and how that cost compares to published lightweight intrusion detection systems. Table~\ref{tab:results} reports exact in-domain and cross-domain figures throughout, and Table~\ref{tab:size_compare} reports the size comparison in Section~\ref{sec:results_sizecompare}; Figures~\ref{fig:indomain_vs_cross}, \ref{fig:shap}, and \ref{fig:fewshot} each visualize a finding not otherwise shown.
\subsection{In-Domain Performance}
\label{sec:results_indomain}

All four architectures achieve closely matched, strong performance on the Edge-IIoTset\cite{ferrag2022edge} test split (Table~\ref{tab:results}, top rows). F1 ranges from 0.971 to 0.972 across all four models, with precision consistently above 0.99 and recall consistently above 0.95. No model separates meaningfully from the others under this condition: the gap between the strongest and weakest in-domain F1 is 0.001, well within the range expected from a single train/test split alone. At the level of individual predictions, DecisionTree~\cite{loh2011classification} and SmallMLP~\cite{taud2017multilayer} disagree on only 5 of 59{,}893 in-domain test instances.

This near-identical performance is itself a methodological outcome, not an incidental one. Under the raw, unbucketed feature representation used in early experimentation, DecisionTree and SmallLSTM~\cite{graves2012long} substantially outperformed SmallMLP and Small1DCNN~\cite{ige2024state} in-domain, a gap traceable to the high-resolution port values both models could exploit. Once ports are reduced to the four coarse buckets described in Section~\ref{sec:datasets}, that gap disappears: all four architectures converge to comparable performance on the same, coarser feature set. The architectures examined here therefore do not differ meaningfully in their capacity to fit Edge-IIoTset; whatever differences emerge in subsequent sections arise from how each architecture behaves once that fit is tested elsewhere.

\subsection{Cross-Domain Performance}
\label{sec:results_crossdomain}

\begin{figure}[ht]
\centering
\includegraphics[width=\columnwidth]{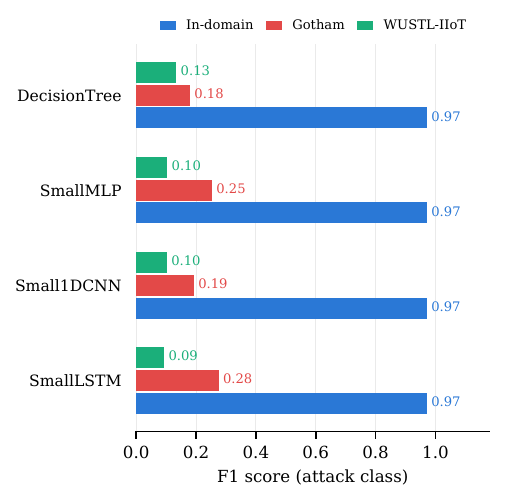}
\caption{In-domain F1 versus cross-domain F1 (natural class distribution) for all four models on both target datasets. Cross-domain performance collapses uniformly relative to in-domain performance, and the collapse is larger on WUSTL-IIoT-2021 than on Gotham for every model.}
\label{fig:indomain_vs_cross}
\end{figure}

Without any retraining, the same four models applied to Gotham dataset~\cite{belarbi2025gotham} and to WUSTL-IIoT-2021~\cite{ahuja2021wustl} show a substantial decline in detection performance relative to the in-domain results of Section~\ref{sec:results_indomain} (Table~\ref{tab:results}, middle and lower rows). Under the natural class distribution, F1 falls from approximately 0.97 in-domain to between 0.18 and 0.28 on Gotham, and to between 0.09 and 0.13 on WUSTL-IIoT-2021 (Fig.~\ref{fig:indomain_vs_cross}). The decline is consistent across all four architectures and both target datasets; under natural-distribution evaluation, no model retains more than 29\% of its in-domain F1 once applied outside its training network.

Under the natural class distribution, the decline is larger on WUSTL-IIoT-2021 than on Gotham for every model. WUSTL-IIoT-2021 differs from Edge-IIoTset in capture

method (flow-level Argus export versus packet-level capture) and lacks TCP flag information entirely (Section~\ref{sec:datasets}), whereas Gotham retains the same packet-level fields as the source domain. The larger drop on the more structurally dissimilar target is consistent with the failure reflecting a loss of transferable signal as the network and capture conditions diverge from the training domain, rather than a fixed quantity independent of how different the target network is.

Comparing balanced and natural evaluation reveals a second, independent finding. On WUSTL-IIoT-2021, balanced-distribution F1 ranges from 0.39 to 0.53, while natural-distribution F1 for the same models falls to 0.09–0.13, a fourfold reduction. The effect is smaller but still present on Gotham (e.g. SmallLSTM: 0.284 balanced versus 0.275 natural). Precision is the metric most affected: under the natural distribution, where attack traffic is rare (7.3\% on WUSTL-IIoT-2021, 10.6\% on Gotham), false positives accumulate against a much larger pool of benign traffic, driving precision to 0.050--0.072 for every model on WUSTL-IIoT-2021
. Recall is comparatively stable between the two evaluation conditions for every model, indicating that the models retain a similar ability to flag attacks once outside their training domain, but the practical cost of that detection, in false alarms relative to true ones, is substantially understated by balanced evaluation alone. This evaluation choice does more than understate the cost of detection: it can reverse which target dataset appears harder. Under balanced evaluation, every model scores higher on WUSTL-IIoT-2021 than on Gotham (e.g. DecisionTree: 0.530 versus 0.171); under natural evaluation, this ordering flips, and every model scores lower on WUSTL-IIoT-2021 than on Gotham (e.g. DecisionTree: 0.134 versus 0.180). Which target network appears to pose the greater generalization challenge therefore depends on the evaluation protocol used, not only on the network itself.

Among the four architectures, SmallLSTM attains the highest cross-domain F1 on Gotham (0.275 natural) but the lowest on WUSTL-IIoT-2021 (0.093 natural); DecisionTree shows the reverse pattern, weakest on Gotham (0.180) but strongest on WUSTL-IIoT-2021 (0.134). No single architecture dominates across both target datasets, indicating that cross-domain ranking is not fixed but depends on which network the model is deployed to.
\begin{table*}[ht] 
\centering
\caption{In-domain and cross-domain detection performance. Cross-domain results are reported under both balanced and natural class distributions.}
\label{tab:results}
\begin{tabular}{llcccc}
\toprule
\textbf{Condition} & \textbf{Model} & \textbf{Acc.} & \textbf{Prec.} & \textbf{Rec.} & \textbf{F1} \\
\midrule
\multirow{4}{*}{In-domain}
& DecisionTree   & 0.972 & 0.993 & 0.951 & 0.972 \\
& SmallMLP       & 0.972 & 0.993 & 0.951 & 0.972 \\
& Small1DCNN     & 0.972 & 0.994 & 0.950 & 0.971 \\
& SmallLSTM      & 0.972 & 0.993 & 0.950 & 0.971 \\
\midrule
\multirow{4}{*}{Cross-domain: Gotham (balanced)}
& DecisionTree   & 0.441 & 0.101 & 0.563 & 0.171 \\
& SmallMLP       & 0.667 & 0.170 & 0.581 & 0.263 \\
& Small1DCNN     & 0.617 & 0.124 & 0.451 & 0.194 \\
& SmallLSTM      & 0.686 & 0.185 & 0.609 & 0.284 \\
\midrule
\multirow{4}{*}{Cross-domain: Gotham (natural)}
& DecisionTree   & 0.449 & 0.107 & 0.567 & 0.180 \\
& SmallMLP       & 0.634 & 0.162 & 0.582 & 0.253 \\
& Small1DCNN     & 0.593 & 0.122 & 0.455 & 0.192 \\
& SmallLSTM      & 0.658 & 0.178 & 0.612 & 0.275 \\
\midrule
\multirow{4}{*}{Cross-domain: WUSTL-IIoT (balanced)}
& DecisionTree   & 0.363 & 0.363 & 0.976 & 0.530 \\
& SmallMLP       & 0.277 & 0.302 & 0.741 & 0.429 \\
& Small1DCNN     & 0.278 & 0.300 & 0.723 & 0.424 \\
& SmallLSTM      & 0.247 & 0.278 & 0.659 & 0.391 \\
\midrule
\multirow{4}{*}{Cross-domain: WUSTL-IIoT (natural)}
& DecisionTree   & 0.078 & 0.072 & 0.976 & 0.134 \\
& SmallMLP       & 0.062 & 0.056 & 0.743 & 0.104 \\
& Small1DCNN     & 0.071 & 0.055 & 0.726 & 0.102 \\
& SmallLSTM      & 0.056 & 0.050 & 0.665 & 0.093 \\
\bottomrule
\end{tabular}
\end{table*}
\subsection{Statistical Validation}
\label{sec:results_validation}

Two checks test whether the results in Sections~\ref{sec:results_indomain} and~\ref{sec:results_crossdomain} reflect a stable property of each model rather than an artifact of one particular train/test split.

Repeating training across five random seeds leaves the central finding unchanged. In-domain F1 is highly stable for every model (standard deviation $\leq 0.002$ for three of the four architectures, $0.002$ for Small1DCNN), confirming that the near-identical in-domain performance reported in Section~\ref{sec:results_indomain} is not a coincidence of one split. Cross-domain F1 (Gotham, balanced distribution) is also stable for three of the four models (DecisionTree: $0.179\pm0.011$; SmallMLP: $0.252\pm0.020$; SmallLSTM: $0.254\pm0.021$), confirming that the collapse reported in Section~\ref{sec:results_crossdomain} holds across seeds rather than reflecting one unfavorable split. Small1DCNN's cross-domain F1 (same condition) is noticeably less stable across seeds ($0.277\pm0.106$), indicating that its cross-domain behavior, while consistently poor, is also the least predictable of the four architectures from one training run to the next.

McNemar's test~\cite{pembury2020effective}, applied pairwise on identical test instances, gives a more cautious picture for in-domain comparisons than the raw F1 values alone might suggest. Despite an in-domain F1 spread of only 0.004 across all four models (Table~\ref{tab:results}), three of the six pairwise comparisons are statistically significant at $p<0.05$: DecisionTree differs significantly from both Small1DCNN and SmallLSTM, and SmallMLP differs significantly from SmallLSTM. The remaining three pairs, including DecisionTree versus SmallMLP, are not significant. This is consistent with what the in-domain test set size implies: with roughly 60{,}000 test instances, even small, consistent differences in per-sample correctness can reach statistical significance without corresponding to a difference of practical consequence. The significant pairs should accordingly be read as evidence that small in-domain differences are not due to chance, not as evidence that any one architecture is meaningfully better suited to the in-domain task than another.

For cross-domain predictions, all six pairwise comparisons are significant at $p<0.05$, including pairs whose absolute F1 values are close, such as SmallMLP versus SmallLSTM (0.253 versus 0.275, natural distribution). Combined with the multi-seed results, this indicates that the differences in cross-domain F1 between architectures, while all of them poor in absolute terms, are systematic rather than incidental: a given architecture's specific degree of cross-domain failure is a reproducible property of that architecture on a given target dataset, not noise around a single common failure rate.
\subsection{Feature Reliance}
\label{sec:results_feature}

\begin{figure*}[t]
\centering
\includegraphics[width=\textwidth]{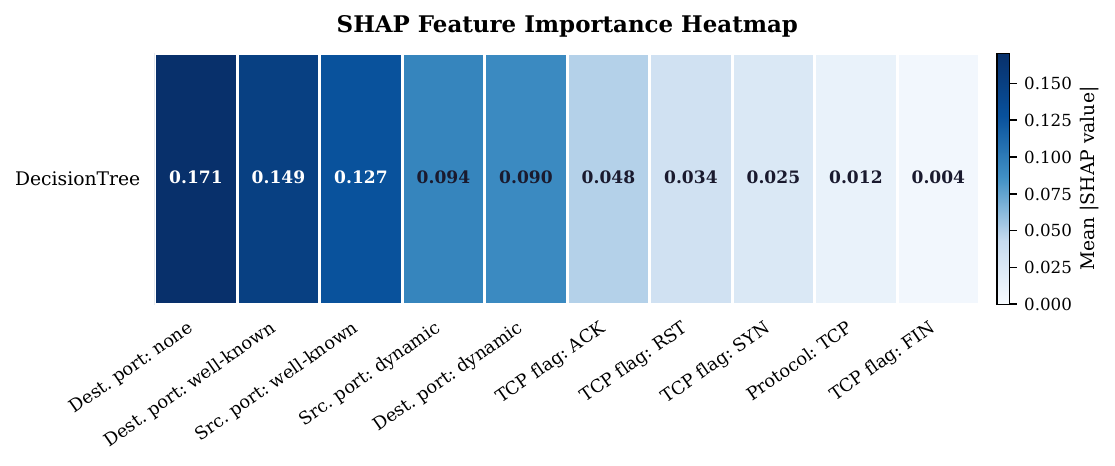}
\caption{Mean absolute SHAP value per feature for DecisionTree, ranked by importance. The ten highest-importance features are shown, of which the top five account for 83\% of total importance across all sixteen features; the remaining six features, including all five zero-importance features, are omitted. Port-bucket features dominate; protocol and TCP flag features contribute comparatively little. The same category-level pattern holds for SmallMLP (Section~\ref{sec:results_feature}).}
\label{fig:shap}
\end{figure*}

SHAP values are computed for both DecisionTree and SmallMLP, the two best in-domain models at full precision (F1 = 0.97155 and 0.97146 respectively, a difference within the multi-seed noise floor reported in Section~\ref{sec:results_validation}), to confirm the port-bucket reliance finding is not an artifact of an arbitrary tie-break (Fig.~\ref{fig:shap}). Port-bucket indicators dominate for both models, though the specific feature relied on most differs: DecisionTree's top feature is \texttt{dst\_port\_none}, while SmallMLP's is \texttt{dst\_port\_dynamic}. Across both models, the five highest-ranked features are entirely port-bucket indicators, accounting for 83\% of total importance for DecisionTree and 73\% for SmallMLP. TCP flags and protocol contribute comparatively little, and five of the sixteen features, \texttt{src\_port\_registered}, \texttt{dst\_port\_registered}, \texttt{protocol\_UDP}, \texttt{protocol\_ICMP}, and \texttt{protocol\_OTHER}, receive zero importance. The model's native Gini-based feature importances agree: individually, \texttt{src\_port\_wellknown} (0.366), \texttt{dst\_port\_wellknown} (0.298), and \texttt{dst\_port\_none} (0.270) account for 93\% of total split importance between them.

This concentration explains the protocol features' negligible contribution. Edge-IIoTset's training traffic is 88.4\% TCP and 11.5\% ICMP, with effectively no UDP; protocol carries almost no discriminative signal despite ICMP comprising 11.5\% of training traffic, indicating that protocol type does not correlate strongly with the attack label in this dataset, rather than reflecting an absence of protocol variation.
Whether this reliance transfers can be tested directly by comparing port-bucket prevalence on attack traffic across all three datasets. \texttt{dst\_port\_none}, DecisionTree's top feature by SHAP value, is present in 40.5\% of Edge-IIoTset's attack traffic, but only 0.42\% of Gotham's and 0.09\% of WUSTL-IIoT-2021's, a 96-fold and 435-fold difference respectively. The features each model has learned to treat as its strongest indicator of an attack are, in both target networks, almost entirely absent from attack traffic, regardless of which specific port-bucket category each model happens to rely on most.
Port-bucket reliance was the explicit target of the ablation in Section~\ref{sec:datasets}: raw port values were replaced with four coarse buckets specifically because high-resolution ports are a documented source of dataset-specific shortcut learning \cite{engelen2021troubleshooting}. Figure~\ref{fig:shap} shows that this reliance was reduced, not removed. The model still depends overwhelmingly on which port bucket a flow falls into, just at a coarser resolution than the original numeric value. The cross-domain collapse reported in Section~\ref{sec:results_crossdomain} therefore cannot be attributed to raw port memorization alone, since that specific mechanism was already constrained; a port-bucket-level dependency of this magnitude is sufficient on its own to explain why performance does not transfer to networks with a different port-bucket distribution.
\subsection{Adversarial Robustness}
\label{sec:results_adversarial}

Robustness to evasion is assessed independently of cross-domain transfer, using the HopSkipJump black-box attack~\cite{chen2020hopskipjumpattack} on 100 in-domain test instances per model (Eq.~\ref{eq:advdrop}); clean accuracy here is computed on this same 100-instance subsample, which is why it differs slightly from the full-test-split in-domain accuracy of 0.972 reported in Table~\ref{tab:results}. All four models start from comparable clean accuracy on this subsample (0.94--0.95) but diverge sharply once perturbed. DecisionTree and Small1DCNN degrade by 0.44--0.45, retaining roughly half their clean accuracy under attack. SmallMLP and SmallLSTM degrade by 0.88, falling to 0.06 accuracy, near the level of guessing a single class on a balanced sample.

This pattern does not track either in-domain performance or cross-domain performance: SmallMLP and SmallLSTM are among the strongest cross-domain performers (Section~\ref{sec:results_crossdomain}) yet the least robust to adversarial perturbation, while DecisionTree, weakest on Gotham, is comparatively robust. Generalization to a different network and robustness to an adversarial input are therefore distinct properties of these architectures, and improving one offers no indication of the other. Unlike the cross-domain and in-domain results above, this evaluation reflects a single 100-sample run without repetition across seeds; the relative ranking between models is consistent with the magnitude of the gap observed, but confidence intervals comparable to those reported elsewhere in this paper are not available for these estimates.

\subsection{Few-Shot Domain Recovery}
\label{sec:results_fewshot}

\begin{figure*}[ht]

\centering
\includegraphics[width=\textwidth]{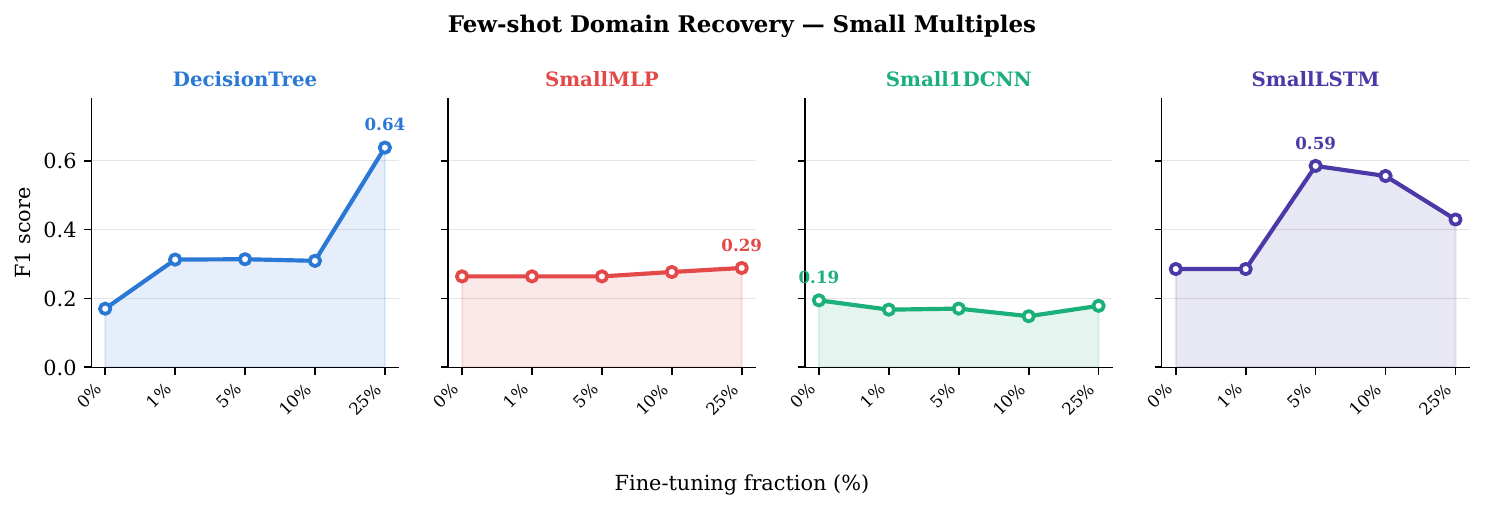}
\caption{F1 on the Gotham evaluation set after fine-tuning each model on increasing fractions of labeled Gotham data. Recovery is architecture-dependent: DecisionTree and SmallLSTM show substantial gains, SmallMLP improves slowly, and Small1DCNN does not improve.}
\label{fig:fewshot}
\end{figure*}

The zero-shot point on each recovery curve (fraction $=0$) is evaluated on the fixed held-out half of the balanced Gotham sample reserved for few-shot evaluation, not the full Gotham sample used in Table~\ref{tab:results}; this is why, for example, DecisionTree's zero-shot F1 of 0.170 here differs slightly from its Gotham-balanced F1 of 0.171 in Table~\ref{tab:results}. Fine-tuning on a small sample of labeled Gotham data, following the procedure in Section~\ref{sec:fewshot}, produces markedly different outcomes across the four architectures (Fig.~\ref{fig:fewshot}). DecisionTree shows no improvement until 25\% of the adaptation pool (14{,}805 samples), at which point F1 nearly quadruples relative to its zero-shot baseline, from 0.170 to 0.638. SmallLSTM improves immediately at 5\% (2{,}961 samples), reaching 0.585, but then declines at 10\% and 25\%, ending at 0.429, below its own 5\% peak. SmallMLP changes little before 10\%, then rises gradually to 0.289 at 25\%, a modest gain over its 0.264 baseline. Small1DCNN does not improve at any fraction tested; its few-shot F1 remains at or below its zero-shot value throughout.

No single relationship between data quantity and recovery holds across all four models. DecisionTree needs a comparatively large sample before any benefit appears; SmallLSTM benefits from a small sample but loses some of that benefit as more data is added; SmallMLP requires data but recovers only partially; Small1DCNN does not recover regardless of sample size. The capacity to close the cross-domain gap through limited local adaptation is therefore a property of the architecture being adapted, not solely a function of how much target-domain data is made available.

\subsection{Efficiency Profiling}
\label{sec:results_efficiency}

All four models remain within the resource bounds expected of edge deployment: model size ranges from 4.7\,KB (SmallMLP) to 7.9\,KB (SmallLSTM), and mean inference latency per sample ranges from 0.11\,ms (SmallMLP) to 0.40\,ms (SmallLSTM). Training cost differs by over two orders of magnitude (425$\times$), from 0.20\,s for DecisionTree to 85.0\,s for SmallLSTM, reflecting the cost of gradient-based training~\cite{bengio2012practical} relative to a single tree fit. None of these differences track the cross-domain or few-shot results of Sections~\ref{sec:results_crossdomain} and~\ref{sec:results_fewshot}: DecisionTree is both the cheapest to train and one of the two architectures that recovers well under few-shot adaptation, while SmallLSTM is the most expensive to train yet performs comparably or better under cross-domain transfer on Gotham; its few-shot recovery, by contrast, peaks early and declines (Section~\ref{sec:results_fewshot}), ending well below DecisionTree's recovered performance. Efficiency and adaptability are independent considerations for model selection in this setting, not substitutes for one another.

\subsection{Comparison to Published Lightweight IDS Footprints}
\label{sec:results_sizecompare}

Table~\ref{tab:size_compare} confirms that all four architectures evaluated here sit comfortably within typical edge-deployment budgets, alongside recent lightweight intrusion detection studies for IoT, IIoT, and resource-constrained edge settings more generally. The comparison is not like-for-like: the studies in Table~\ref{tab:size_compare} achieve compactness through compression, quantization, or distillation applied to a richer native feature set, while the smaller size reported here follows directly from the minimal 16-dimensional common schema used throughout this study (Section~\ref{sec:datasets}), not from compressing an otherwise larger model. One comparison point, DNN-KDQ \cite{umar2025dnnkdq}, targets edge deployment generally and is benchmarked on an enterprise network dataset rather than an IoT or IIoT capture; it is included as a edge-deployment efficiency reference rather than a domain-matched comparison. All models in Table~\ref{tab:size_compare}, including those evaluated here, fall well below the 250--500\,KB deployment budget commonly cited in the broader TinyML literature \cite{tinymlsurvey2025}; the differences within the table (4.7--39.9\,KB) are accordingly a confirmation that all of these approaches are genuinely lightweight by that standard, rather than a finding with independent deployment consequences.

\begin{table*}[ht]
\centering
\caption{Model size comparison against recently published lightweight intrusion detection systems.}
\label{tab:size_compare}
\begin{tabular}{lcc}
\toprule
\textbf{Model / Study} & \textbf{Size} & \textbf{Latency} \\
\midrule
DecisionTree (this work)     & 5.7\,KB  & 0.16\,ms \\
SmallMLP (this work)         & 4.7\,KB  & 0.11\,ms \\
Small1DCNN (this work)       & 4.8\,KB  & 0.21\,ms \\
SmallLSTM (this work)        & 7.9\,KB  & 0.40\,ms \\
\midrule
TCN \cite{tcn2026edge}                  & 16.3\,KB & -- \\
TinyML FNN \cite{lundqvist2025tinyml}   & 31\,KB   & -- \\
Dynamic-Quant.\ IDS \cite{discoveriot2025} & $<$32\,KB & -- \\
BiGRU-MHA-LSTM \cite{tcn2026edge}       & 39.9\,KB & -- \\
DNN-KDQ \cite{umar2025dnnkdq}           & 20.2\,KB & 0.07\,ms \\
\bottomrule
\end{tabular}
\end{table*}
\section{Discussion}
\label{sec:discussion}

The results in Section~\ref{sec:results} support a specific, narrower claim than "lightweight models do not generalize." Generalization fails consistently and substantially across two independent IIoT datasets, but the mechanism behind that failure is now partially understood, and the failure is not uniform across architectures once adaptation is introduced.

\subsection{Coarsening Features Relocates Shortcut Reliance, It Does Not Remove It}

The port-bucketing ablation (Section~\ref{sec:datasets}) was designed to test whether raw port memorization, a documented shortcut in network intrusion detection \cite{engelen2021troubleshooting}, explains the cross-domain collapse. It does not. DecisionTree's feature importances and SHAP values (Section~\ref{sec:results_feature}) show that the model still depends overwhelmingly on which coarse port-bucket category a flow falls into, with three categories accounting for 93\% of split importance; SmallMLP, an architecturally unrelated model tied with DecisionTree in-domain, independently shows the same category-level dependence, indicating this is not an artifact of one model's particular fit. Coarsening removed the original, fine-grained shortcut while leaving a coarser version of the same mechanism intact. Section~\ref{sec:results_feature} shows this is sufficient to explain the collapse: the port-bucket feature DecisionTree relies on most heavily by SHAP value occurs at 96 times the rate in source-domain attack traffic compared to Gotham, and 435 times the rate compared to WUSTL-IIoT-2021. This indicates that reducing feature resolution is not, by itself, a reliable defense against shortcut learning in intrusion detection; the shortcut persists at whatever resolution the feature representation still preserves, and transfers only as well as that resolution's distribution does.

\subsection{Balanced Evaluation Overstates Deployment-Relevant Performance}

The natural-distribution results (Section~\ref{sec:results_crossdomain}) show that balanced-class evaluation, the convention in most prior cross-dataset studies, substantially overstates performance under conditions a deployed system would actually face. The effect is largest on WUSTL-IIoT-2021, where natural-distribution F1 is roughly one quarter of its balanced-distribution value, driven primarily by a collapse in precision as false positives accumulate against a much larger pool of benign traffic. This suggests that F1 or accuracy figures reported under artificial class balance should not be read as estimates of operational performance. The effect is not limited to overstating performance on a single target: balanced evaluation reverses which of the two target datasets appears more difficult relative to natural evaluation (Section~\ref{sec:results_crossdomain}), indicating that comparative claims about which deployment environment poses the greater generalization challenge are themselves protocol-dependent, not solely a property of the environments being compared.
For context, Ferrag et al. report 99.54\% accuracy for binary classification on Edge-IIoTset using the dataset's full native feature set \cite{ferrag2022edge}; the in-domain accuracy reported here (97.2\%, Table~\ref{tab:results}) reflects the cost of restricting the feature set to sixteen dimensions deliberately chosen for cross-dataset compatibility (Section~\ref{sec:datasets}); the 2.3-point difference is the price paid for a schema designed to generalize, not evidence of a weaker model. No equivalent cross-dataset baseline exists for direct comparison on these three datasets, which is the gap this paper addresses.

\subsection{Adaptability Is Architecture-Dependent, Not Data-Dependent}
\label{sec:disc_adapt}
The few-shot results (Section~\ref{sec:results_fewshot}) show that no single rule relates the amount of target-domain data available to the degree of recovery obtained. DecisionTree requires a comparatively large adaptation sample before any benefit appears, then recovers substantially. SmallLSTM recovers with a much smaller sample, but its performance declines as more adaptation data is added rather than continuing to improve; a plausible contributor is the per-batch class-weight recomputation used during fine-tuning (Eq.~\ref{eq:loss}), which changes as the adaptation sample's class balance shifts with size, though confirming this would require repeating the experiment under fixed class weights. SmallMLP improves only modestly even with the largest sample tested, and Small1DCNN does not improve at any sample size. A substantial part of DecisionTree's recovery may reflect this procedural difference rather than architecture alone, since it is refit on its full original training set plus the new sample while the neural models are fine-tuned on the new sample only; this study does not include a DecisionTree-refit-on-small-sample-alone condition that would isolate the two factors, and the magnitude of DecisionTree's recovery should accordingly be read as a property of the update procedure used, not of decision trees as an architecture class. The architecture-dependence claim rests more securely on the comparison between SmallLSTM and Small1DCNN, which are fine-tuned under identical procedures yet diverge sharply in outcome (Section~\ref{sec:results_fewshot}); this pair, not the DecisionTree comparison, is the cleanest evidence for Contribution 4. This comparison should still be read with one caveat: SmallLSTM processes the 16-dimensional feature vector as an artificial sequence of scalars rather than a natural temporal signal (Section~\ref{sec:methodology}), so its divergence from Small1DCNN may partly reflect this specific parameterization choice rather than a general property of recurrent architectures.
\subsection{Generalization, Robustness, and Efficiency Are Independent Axes}

Three properties examined in this paper, cross-domain generalization, adversarial robustness, and computational efficiency, do not track one another. DecisionTree is the cheapest to train, among the more adversarially robust models, and one of the two architectures that recovers well under few-shot adaptation; SmallLSTM is the most expensive to train yet performs comparably or better on cross-domain transfer while being among the least adversarially robust. A model that is favorable on one of these axes offers no guarantee on the others, and a deployment decision that optimizes for one without examining the rest risks an unexamined weakness on another. This independence is not absolute: on WUSTL-IIoT-2021 specifically, DecisionTree is both the strongest cross-domain performer and among the more adversarially robust models, an alignment rather than an independence; the broader pattern holds because SmallMLP and SmallLSTM break it in the opposite direction on Gotham.

\subsection{Practical Implications for Model Selection}

No single architecture is preferable on every axis examined here. If deployment can budget for a modest labeled sample from the target network, DecisionTree offers the strongest combination of low training cost and substantial few-shot recovery, though the confound noted in Section~\ref{sec:disc_adapt} regarding its update procedure should be kept in mind. If zero-shot performance on an unfamiliar network is the priority and a known adversarial robustness cost is acceptable, SmallLSTM performs best of the four on Gotham, though not on WUSTL-IIoT-2021. Small1DCNN is not recommended under any condition tested here: it does not generalize well zero-shot, does not recover with adaptation, and offers no efficiency advantage over the other three architectures. These recommendations are specific to the two target networks tested and should be revisited for deployment targets that differ structurally from both.

\subsection{Limitations}

Five limitations bound the scope of these findings. First, the common feature schema is restricted to fields available across all three datasets; WUSTL-IIoT-2021 in particular contributes no TCP flag information, so its cross-domain results reflect only the port-bucket and protocol portion of the schema. Second, Edge-IIoTset's training traffic is 88.4\% TCP and 11.5\% ICMP, with effectively no UDP traffic, limiting how much protocol diversity the source domain offers for a model to learn from. Third, adversarial robustness is reported from a single 100-sample evaluation per model without repetition across seeds, in contrast to the multi-seed and significance testing applied to the cross-domain results; the relative ordering between models should be read with correspondingly lower confidence than the rest of this paper's findings. Fourth, few-shot recovery is evaluated on Gotham only; Section~\ref{sec:results_crossdomain} shows that architecture ranking under zero-shot transfer differs between Gotham and WUSTL-IIoT-2021, so whether the same architecture-dependent recovery pattern observed here also holds on WUSTL-IIoT-2021 is untested. Fifth, the explainability analysis in Section~\ref{sec:results_feature} is conducted on DecisionTree and SmallMLP; both rely predominantly on port-bucket categories, though on different specific features, strengthening the generality of the port-shortcut finding. Whether Small1DCNN and SmallLSTM rely on the same category of feature is not directly verified, and is inferred only indirectly from their shared cross-domain collapse pattern. Sixth, WUSTL-IIoT-2021's larger cross-domain collapse relative to Gotham is attributed to general structural dissimilarity, but its zero-filled TCP flag features are a specific, isolable contributor not separated from other sources of dissimilarity in this study; a controlled ablation removing flag features from all three datasets uniformly would be needed to attribute the larger collapse specifically to this factor rather than to capture-method or protocol differences more broadly. Seventh, the class-weight recomputation hypothesis offered for SmallLSTM's non-monotonic few-shot recovery (Section~\ref{sec:disc_adapt}) was not tested directly; confirming it would require repeating the few-shot experiment with fixed class weights across all fractions.
\section{Conclusion}
\label{sec:conclusion}

This paper tested whether lightweight intrusion detection models that perform well on the network they were trained on retain that performance on a different IIoT network. Across two independent target datasets, they do not: detection performance collapses substantially and consistently, a result stable across random seeds and statistically significant. This failure persists even after a feature-resolution ablation specifically designed to remove the most obvious explanation, raw port memorization, indicating that the same shortcut mechanism operates at coarser feature resolutions and is not eliminated by reducing granularity alone. Evaluating under each target network's natural, imbalanced class distribution rather than the artificially balanced convention common in prior work reveals a further, independent problem: balanced evaluation not only overstates performance but can reverse which deployment target appears more difficult. Recovery through small-scale fine-tuning on target-domain data is possible for some architectures but not others, indicating that adaptability is a property of model design rather than of lightweight models as a class. Taken together, these findings indicate that within-domain accuracy is not a reliable basis for judging deployment readiness, and that cross-network evaluation, under realistic class distributions, should be treated as a standard requirement for lightweight IIoT intrusion detection research rather than an optional addition.

\section*{CRediT authorship contribution statement}

\textbf{MD Azizul Hakim}: Conceptualization, Methodology, Software, Formal
analysis, Visualization, Writing -- original draft. \textbf{Md Shihab
Uddin}: Investigation, Formal analysis, Visualization, Writing -- Review and Editing. \textbf{Talha Ibne Anich}: Investigation, Formal analysis,
Visualization, Writing -- Review and Editing.

\section*{Declaration of generative AI and AI-assisted technologies in the
manuscript preparation process}
During the preparation of this work the authors used Sonnet 4.6 in
order to paraphrase some portion of text and to check the grammar of
the manuscript. After using this tool/service, the authors reviewed and
edited the content as needed and take full responsibility for the content
of the published article

\section*{Declaration of competing interest}
The authors declare that they have no known competing financial interests or personal relationships that could have appeared to
influence the work reported in this paper.

\section*{Acknowledgements}
This research did not receive any specific grant from funding agencies in the public, commercial, or not-for-profit sectors. All costs associated with this study, including computational and data resources, were borne by the authors.

\section*{Data availability}
Edge-IIoTset~\cite{ferrag2022edge}, Gotham 2025~\cite{belarbi2025gotham}, and WUSTL-IIoT-2021~\cite{ahuja2021wustl} are publicly available from their original sources. Code to reproduce the experiments in this study is available \href{https://github.com/logicsame/Cross-Domain-Generalization-Failure-in-Lightweight-Intrusion-Detection-Models-for-IIoT-Networks}{here}.

\bibliographystyle{unsrtnat}
\bibliography{IEEEbiography}

\end{document}